\documentclass[aps,prl,twocolumn,showpacs,groupedaddress,amsmath,amssymb]{revtex4}
\usepackage{graphicx}
\usepackage{color}
\usepackage{bm}
\usepackage{latexsym}

\begin{document}

\title{Light scattering in pseudo-passive media with uniformly balanced gain and loss}
\author{A. Basiri$^{1}$, I. Vitebskiy$^{2}$, T. Kottos$^{1}$}
\affiliation{$^1$Department of Physics, Wesleyan University, Middletown, CT-06459, USA}
\affiliation{$^2$Air Force Research Laboratory, Sensors Directorate, Wright Patterson AFB, OH 45433 USA}
\date{\today }

\begin{abstract}
We introduce a class of metamaterials with uniformly balanced gain and loss associated with complex permittivity and permeability 
constants. The refractive index of such a balanced pseudo-passive metamaterial is real. An unbounded uniform pseudo-passive 
medium has transport characteristics similar to those of its truly passive and lossless counterpart with the same real refractive index. 
However, bounded pseudo-passive samples show some unexpected scattering features which can be further emphasized by including 
such elements in a photonic structure.
\end{abstract}

\pacs{42.82.Et, 42.25.Bs, 11.30.Er}
\maketitle

In the last decade we have witnessed unprecedented advancements in realizing artificial materials which are specifically designed to 
exhibit features not found in nature. In the electromagnetic domain, such metamaterials are using their structural composition in order
to obtain access of all four quadrants in the real $\epsilon-\mu$ plane. Various exotic phenomena ranging from negative refraction 
to electromagnetic cloaking and super-lensing have been actively pursued \cite{Y87,JMW95,BKR97,K03,V68,P00,SSS01,SCCYSDK05}. 
While these opportunities have been extraordinary, optical materials exhibiting exotic values of permittivity $\varepsilon$\ and/or 
permeability $\mu$ are often prohibitively lossy. This is especially true for composite optical meta-materials. A natural solution to the 
problem is to add a gain component and, thereby, to offset the losses. A pathway to achieve this goal has been recently proposed in 
Ref. \cite{MGCM08}, and subsequently explored in a number of publications \cite{GSDMVASC09,RMGCSK10,CJHYWJLWX14,RBMOCP12,
FWMWZ14,KGM08,L09,L10,CGS11,LPRS13,RMBNOCP13,HMHCK14,CSGAE13,POLMGLFNBY14,LFLVEK14,NBRMCK14,RLKKKV12,RCKVK12,
LREKCC11}. This proposal capitalizes on the notion of Parity-Time (${\cal PT}$) symmetry by utilizing balanced gain and loss elements 
which are judiciously distributed in space such that the complex index of refraction $n(r)$ satisfies the relation $n(r)=n^*(-r)$. This 
research line gave many intriguing transport properties like double refraction, unidirectional invisibility, asymmetric transport, CPA/
Lasers etc.

An alternative approach is to compensate the losses with gain while preserving the uniformity of the medium. This can be done 
simply by doping the lossy medium with active elements (dies), so that the doped material would have a real refractive index
$n(r)=\sqrt{\varepsilon\mu}=n^{\ast}(r).$ The question, however, is whether such a balanced loss/gain medium with real and 
uniform refractive index will automatically behave as a regular passive lossless medium. This is certainly true if the permeability 
$\mu$ is real, in which case the negative contribution to $\varepsilon^{\prime\prime}$ from gain will simply offset the positive 
contribution from absorption \cite{gain1,gain2,gain3}. 

\begin{figure}[h]
\includegraphics[width=8cm,height=5cm,clip]{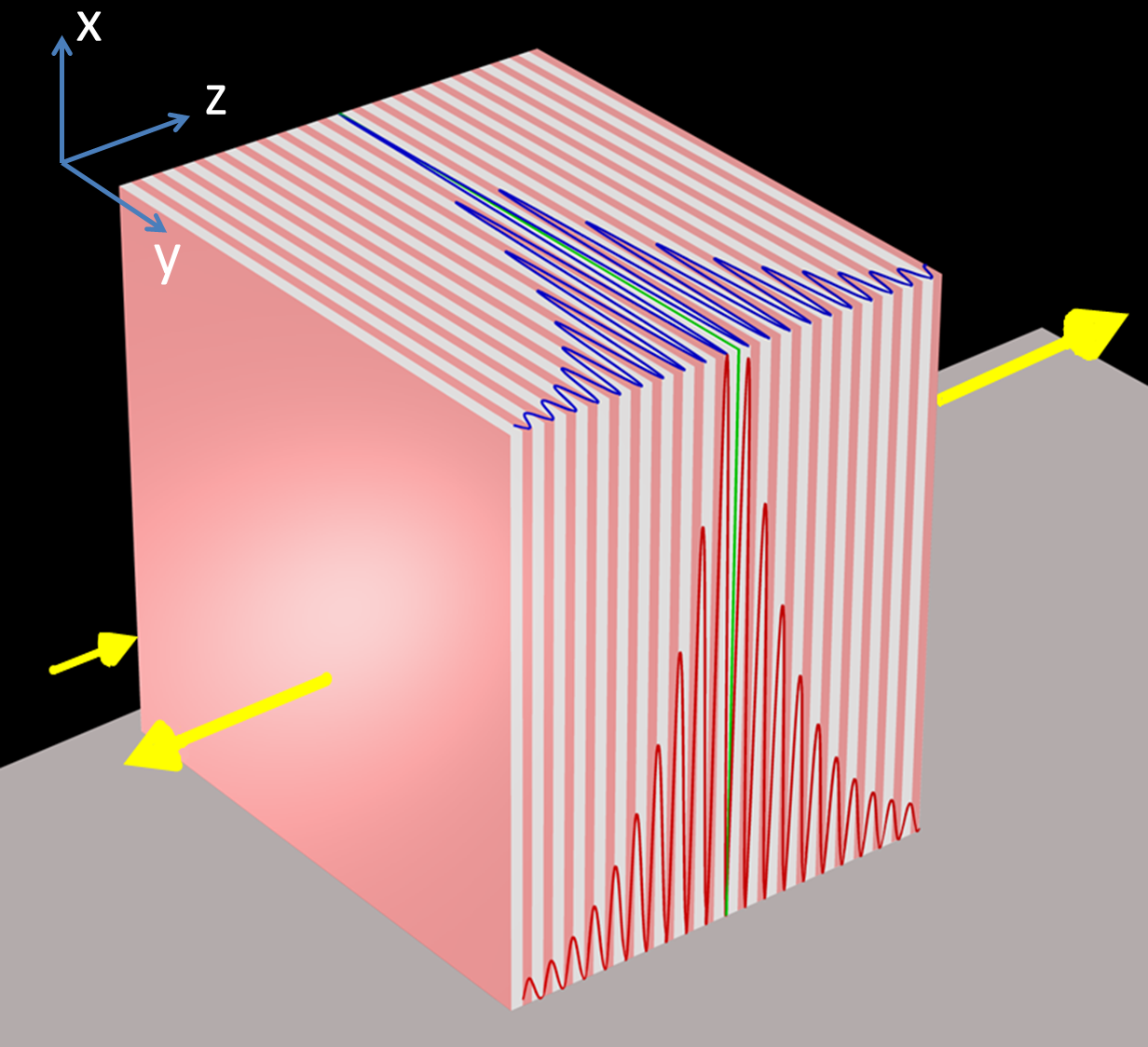}
 \caption{
   \label{fig1}
(color online) The electric $E(z)$ (y-z plane) and the magnetic $H(z)$ (x-z plane) field distribution of a resonance localized mode 
$\lambda_r=1.256\mu m$ for a Bragg grating with an embedded PP-defect (green). The grating consists of 20 lossless bilayers (orange 
and white) at each side of the defect. Their index of refraction is $n_1=1.45$ and $n_2=1.755$ and their width is $d_1=0.2167\mu m$ 
and $d_2=0.179\mu m$. The pseudo-passive (PP) defect layer has index of refraction $n_p=3.48$ corresponding to constituents 
$\epsilon_p=3.48-i0.01, \mu_p=\epsilon_p^*$ while its width is $d_p=0.0542\mu m$. 
}
\end{figure}

In this Letter we will concern with a different scenario where both $\varepsilon$ and $\mu$ are complex i.e. $\varepsilon=
\varepsilon^{\prime}+i\varepsilon^{\prime\prime},\ \ \mu=\mu^{\prime}+i\mu^{\prime\prime}$. A typical example is plasmonic 
metamaterials where the effective permeability is associated with the electric current induced in tiny ring resonators by 
oscillating magnetic fields. If both $\varepsilon$\ and $\mu$ are complex, a real-value refractive index $n$ can only be realized 
if $\varepsilon=\rho^2 \mu^*$ (where $\rho$ is real), which is quite possible in media with gain. In this case $n=\rho |\mu|$ 
and thus one expects that an unbounded medium consisting of such material will support traveling waves as its lossless passive 
counterpart with the same index of refraction. We will therefore refer to media with real refractive index $n$ and complex 
$\varepsilon^{\prime\prime}$ and $\mu^{\prime\prime}$ as pseudo-passive (PP) media. Surprisingly, once a pseudo-passive 
medium is confined in space, its scattering properties are completely different than the ones of its passive counterpart. The 
difference becomes even more striking when a pseudo-passive material is a component of a photonic structure. For example,
the Fabry-Perrot resonances of a Bragg grating that contains pseudo-passive composite layers display sub- or super unitary 
transmittance at the band-edges. Remarkably, under different circumstances, the same pseudo-passive material can act as a 
perfect absorber, or it can trigger lasing instability.

The scattering properties of a pseudo-passive structure are understood better by considering monochromatic wave propagation 
in an one-dimensional scattering set-up. In this case, the steady state (TE) electric field $E(z)$ is scalar and satisfies the Helmholtz 
equation
\begin{equation}
\label{Helm}
\frac {d^2 E(z)}{dz^2} +n^2(z) \left(\frac{\omega}{c}\right)^2 E(z)=0
\end{equation}
where $\omega$ is the frequency of the field and $c$ is the speed of light in vacuum while the spatially dependent index of 
refraction $n(z)=\sqrt{\varepsilon(z) \mu(z)}$ is considered real for any position $z$. In the case that the pseudo-passive
medium is extended over the whole space, Eq. (\ref{Helm}) admits wave solutions of the form $E(z)=E^{+} \exp(iknz) + E^{-}
\exp(-iknz)$ where $k=\omega/c$ is the free space wavevector. These solutions are identical to the ones found at an infinite 
lossless passive medium with index of refraction $n$.  

These similarities between a PP medium and a lossless passive medium cease to exist once we turn to the analysis of the associated
scattering problem involving bounded structures. For simplicity we assume that the domain which contains the pseudo-passive 
structure extends in the interval $0<z<L$. We further assume that the scatterer is embedded in a homogeneous medium with 
uniform index of refraction $n_0$ which, for simplicity, we consider it to be equal to unity i.e. $n_0=1$. The solution of Eq. (\ref{Helm}) 
at the left of the scattering sample $z<0$ is $E_L=E_L^{+} \exp(ikz)+E_L^{-}\exp(-ikz)$ while on its right $z>L$ is $E_R=E_R^{+} 
\exp(ikz)+E_R^{-}\exp(-ikz)$. We can relate the amplitudes of forward and backward propagating waves on the left of the scattering 
domain with the amplitudes on its right via the transfer matrix ${\cal M}$ 
\begin{eqnarray}
\begin{pmatrix}
 E_{R}^{+} \\
 E_{R}^{-}
\end{pmatrix}
&=&{\cal M}
\begin{pmatrix}
  E_{L}^{+} \\
 E_{L}^{-}
\end{pmatrix}.\,\label{Mp1}
\end{eqnarray}
In case of composite structures the transfer matrix (TM) ${\cal M}$ is a product of transfer matrices ${\cal M}_n$ associated with 
each individual composite element i.e. ${\cal M}=\prod_n {\cal M}_n$. The individual TM's are evaluated by 
imposing the appropriate boundary conditions at the interface between consequent layers $n$ and $n+1$ of the structure. For a TE 
mode we have that the field itself is continuous while its spatial derivative satisfies the relation:
\begin{equation}
\label{spatder}
\left [
{1\over \mu_n} \left( \partial E_n(z) \over \partial z\right) = {1\over \mu_{n+1}} \left( \partial E_{n+1}(z) \over 
\partial z\right) \right ]_{\rm interface}.
\end{equation}

The transmission $t$ and reflection $r$ amplitudes for left and right incident waves are defined as $t_{L}\equiv E_{R}^{+}/E_{L}^{+},
r_{L}\equiv E_{L}^{-}/E_{L}^{+}$ and $t_R\equiv E_{L}^{-}/E_{R}^{-}, r_R\equiv E_{R}^{+}/E_{R}^{-}$. They are written in terms of the 
TM elements as 
\begin{equation}
\label{TR}
t_L={\det {\cal M}\over {\cal M}_{22}},  t_R={1\over {\cal M}_{22}},
r_L=-{{\cal M}_{21}\over {\cal M}_{22}}, r_R={{\cal M}_{12}\over {\cal M}_{22}}. 
\end{equation}
Furthermore, we can show that $\det {\cal M}=1$ and thus $t_L=t_R=t$. The associated transmittance ${\cal T}$ and reflectances 
${\cal R}_{L,R}$ are ${\cal T}=|t|^2$ and ${\cal R}_{L,R}=|r_{L,R}|^2$. 

An alternative formulation of the scattering process is provided by the scattering matrix $S$ which connects incoming and outgoing 
wave amplitudes i.e.
\begin{equation}
\label{scat1}
\left(
\begin{array}{c}
E_L^{-}\\
E_R^{+}
\end{array}
\right)=
S\left(\begin{array}{c}
E_L^{+}\\
E_R^{-}
\end{array}
\right)
\end{equation}
The scattering matrix can be written in terms of the ${\cal M}$-matrix elements as
\begin{equation}
\label{scat2}
S=\left(
\begin{array}{cc}
r_L  &t_R\\
t_L  &r_R
\end{array}
\right)=
{1\over {\cal M}_{22}}\left(
\begin{array}{cc}
-{\cal M}_{21}  &1\\
det({\cal M})  &{\cal M}_{12}
\end{array}
\right)
\end{equation}
and is useful for the theoretical analysis of lasing instabilities and perfect absorption. Below we provide some examples 
which illustrate the anomalous scattering properties of a bounded structure that includes a PP-medium.

We start our analysis with the investigation of the transport properties of a single PP-layer. The associated TM that describes 
the scattering process is
\begin{eqnarray}
\label{Mp1}
{\cal M}&=&D(L;n_0)^{-1}
K({n_p\over n_0};{\mu_0\over \mu_p})
D(L;n_p)
K({n_0\over n_p};{\mu_p\over \mu_0})
\end{eqnarray}
where the matrices $D$ and $K$ are defined as
\begin{eqnarray}
\label{Mp2}
K({n_l\over n_m};{\mu_m\over \mu_l})&=&
\frac{1}{2}
\begin{pmatrix}
 1+({n_l\over n_m})({\mu_m\over\mu_l})& 1-({n_l\over n_m})({\mu_m\over\mu_l})  \\
 1-({n_l\over n_m})({\mu_m\over\mu_l})& 1+({n_l\over n_m})({\mu_m\over\mu_l})
\end{pmatrix} \,\,\,\,\,\, \nonumber\\
D(L;n)&=&\begin{pmatrix}
 \exp(iknL) & 0  \\
  0 &\exp(-iknL)
\end{pmatrix}
\end{eqnarray}

Using Eq. (\ref{TR}) we evaluate the transmission and reflection amplitudes of a single PP-layer:
\begin{eqnarray}
t=&\frac{2\exp[-ikL]}{2\cos[k n_p L]-i(y_p+\frac{1}{y_p})\sin[k n_p L]}&\\
r= &r_L=\frac{i(y_p-\frac{1}{y_p}) \sin(k n_p L)}{2\cos[k n_p L]-i(y_p+\frac{1}{y_p})\sin[k n_p L]};& \, r_R=r \exp(-2ikL)\nonumber
\end{eqnarray}
The subindex $p$ indicates that we refer to the constituents of the PP-layer and $y_p\equiv \sqrt{\frac{\epsilon_p}{\mu_p}}$ is 
its complex admittance. 

The Fabry-Perrot resonances are defined by the requirements $T=1$ and  $R=0$ and occur at frequencies $\omega_{\rm FP}=
k_{\rm FP}c$ for which $\sin(k_{\rm FP} n_p L)=0$. This implies perfect resonant transmission with no losses and no gain, 
{\it regardless} of the value of $y_p$ and {\it despite} the fact that the constituents are complex. Let us now calculate the transport 
characteristics of the PP-layer at $\omega\neq \omega_{\rm FP}$. For example, at the middle between two consecutive Fabry-Perrot 
resonances, i.e. when $\cos[k n_pL]=0$, we have:
\begin{eqnarray}
{\cal T}&=\left\vert \frac{2}{y_p+1/y_p}\right\vert ^{2},\,{\cal R}&= \left\vert \frac{\left(y_p-1/y_p\right)}{\left(y_p+1/y_p\right)
}\right\vert ^{2},\\
{\cal A}  & \equiv 1-{\cal T}-{\cal R}&=\frac{2\left(y_p-y_p^{\ast}\right)  ^{2}}{\left\vert y_p^{2}+1\right\vert ^{2}}=-8 \frac{(Im[y_p])^{2}}
{|y_p^2+1|^2}<0\nonumber
\label{eq5}
\end{eqnarray}
Remarkably, the off-resonance value of ${\cal A}$ is always negative and independent of the slab thickness $L$ indicating that its origin 
is associated with surface scattering rather than bulk scattering precesses. In fact, it is straightforward to show that a PP-layer will result 
in amplification ${\cal A}<0$, for any frequency $\omega$, irrespective of the sign of $\epsilon^{\prime\prime}$. 

Next we consider the transport properties of a lossless Bragg grating (BG) with one PP-defect layer. We assume, without loss of generality, 
that $\epsilon^{\prime\prime}<0$. Before going on with our analysis we recall that a lossless passive defect layer supports a resonant 
defect mode with a frequency $\omega_r$ lying inside the photonic band gap of the BG. This resonance mode is localized in the vicinity 
of the defect layer and decays exponentially away from the defect. Near $\omega_r$, the entire composite structure displays a strong 
resonant transmission with ${\mathcal T}=1$ (and thus ${\mathcal R}=0$) due to the excitation of the localized mode.

\begin{figure}[h]
\includegraphics[width=1.0\columnwidth,keepaspectratio,clip]{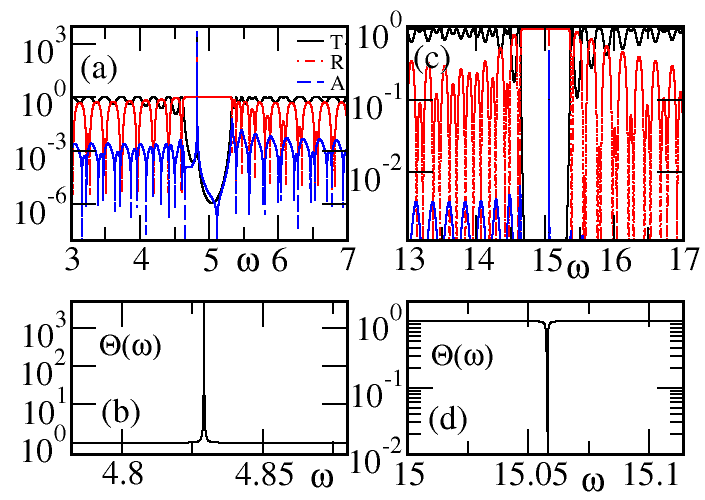}
 \caption{
   \label{fig2}
(color online) (a) The transmittance ${\cal T}(\omega)$, reflectance ${\cal R}(\omega)$ and absorption ${\cal A}(\omega)$ 
for the set-up of Fig. \ref{fig1} ($\epsilon^{\prime \prime}<0$). A super-unitary transmittance and reflectance is evident 
in the frequency domain around the resonance frequency. (b) The amplification/absorption coefficient $\Theta$ versus frequency
has a singular point at resonance frequency which signifies a lasing action. (c) The same set-up as in Fig. \ref{fig1} for the 
case of CPA frequency. In this case the electric field as a minimum at the defect layer while the magentic field has a maximum
(d) The amplification/absorption coefficient $\Theta$ versus frequency shows a zero at the CPA frequency.
}
\end{figure}

A completely different scenario emerges when the defect layer consists of a PP-medium. We find that the resonance localized mode 
has transmittance and reflectance values which are larger (smaller) that unity in distinct contrast to the case of lossless defect layers. 
The super (sub)-unitary transmission/reflection is related with the distribution of the electro-magnetic field around the PP-defect. 
Let us consider the case for which the electric field $E(z)$ takes its maximum value (anti-nodal point) at the position of the defect 
layer (see Fig. \ref{fig1}). Obviously the associated magnetic field $H(z)$ will be having a nodal point at the defect layer. Then, an 
estimation of the absorbed energy, due to the PP-layer, gives ${\mathcal A}\approx |E_d|^2 {\mathcal I}m(\epsilon^{\prime \prime}) 
<0$ and thus ${\cal T + R}=1-{\cal A}\gg 1$. The transmission and reflection spectrum are shown in Fig. \ref{fig2}a and confirm 
the previous expectations. 

The opposite scenario is observed in the case that the magnetic field $H(z)$ has a maximum in the domain around the defect layer. 
In this case ${\cal T + R}=1-{\cal A}\ll 1$ (we recall that ${\cal A}\approx |H_d|^2{\mathcal I}m(\mu^{\prime \prime}) >0$) and 
thus we have attenuation of the incident light.

The amplification/attenuation mechanism is better investigated by introducing the overall amplification/absorption coefficient $\Theta 
(\epsilon^{\prime \prime},\omega)$ defined as the ratio of the total intensity of outgoing to incoming waves:
\begin{equation}
\Theta(\epsilon^{\prime\prime},\omega)\equiv\frac{|{ E_{L}^{-}} |^2+| E_{R}^{+}|^{2}}{|  E_{L}^{+} |^{2}+| E_{R}^{-}|^{2}}
\label{theta}
\end{equation} 
The above expression can be written in the following form
\begin{equation}
\label{theta2}
\Theta(\epsilon^{\prime\prime},\omega)=\frac{\left|1+{E_R^{-}\over E_L^{+}}{\cal M}_{12}\right|^2+\left|{E_R^{-}\over E_L^{+}}-
{\cal M}_{21}\right|^2}{\left|{\cal M}_{22}\right|^2\left(1+\left|{E_R^{-}\over E_L^{+}}\right|^2\right)}
\end{equation}
where we have used the fact that $\det {\cal M}=1$. The case $\Theta(\epsilon^{\prime\prime},\omega) >1$ indicates that an overall 
amplification has been achieved at the system. The opposite limit of $\Theta(\epsilon^{\prime\prime},\omega) <1$ corresponds to 
attenuation. The two extreme cases of $\Theta(\epsilon^{\prime\prime}_L,\omega_L)\rightarrow \infty$ and $\Theta(\epsilon^{\prime
\prime}_{\rm CPA}, \omega_{\rm CPA})\rightarrow 0$ indicate that the system has reached a lasing instability or behaves as a coherent 
perfect absorber (CPA) respectively. The condition for the former case is ${\cal M}_{22}(\epsilon^{\prime\prime},\omega)=0$ which 
is satisfied for some values $\epsilon^{\prime \prime}=\epsilon^{\prime\prime}_L$ and $\omega=\omega_L$. In fact, the complex 
zeros of ${\cal M}_{22}$ correspond to the poles of  the scattering matrix $S$, see Eq. (\ref{scat2}). If $\epsilon^{\prime\prime}=0$ 
they lie at the lower part of the complex-$\omega$ plane due to causality. As $|\epsilon^{\prime\prime}|$ increases the poles move 
towards the real axis and at a critical value $\epsilon^{\prime\prime}=\epsilon^{\prime\prime}_L$ one of these poles becomes real 
$\omega=\omega_L$, thus signifying the transition to a lasing action. We have confirmed in Fig. \ref{fig2}b that the singularity in 
$\Theta(\epsilon^{\prime\prime}=\epsilon^{\prime\prime}_L,\omega=\omega_L)\rightarrow \infty$ corresponds to the lasing point 
as it is calculated from the analysis of the poles of the associated scattering matrix.

The other limiting case of CPA  is achieved when the two terms in the numerator of Eq. (\ref{theta2}) become simultaneously  zero 
i.e. when $E_R^{-}= E_L^{+}{\cal M}_{21}$ and $E_R^{-}= -E_L^{+}/{\cal M}_{12}$. These two relations are simultaneously satisfied 
when ${\cal M}_{21}{\cal M}_{12}+1=0\rightarrow {\cal M}_{11}{\cal M}_{22}=0$ (recall that $\det {\cal M}=1$). The CPA frequency 
$\omega=\omega_{\rm CPA}$ is evaluated from the condition ${\cal M}_{11}(\omega_{\rm CPA})=0$ (the frequencies for which 
${\cal M}_{22}(\omega)=0$ are excluded since they correspond to lasing action). It should be stressed that a CPA requires coherent
incident fields which satisfy an appropriate phase and amplitude relationship $E_R^{-}= E_L^{+}{\cal M}_{21}= -E_L^{+}/{\cal M}_{12}$. 
An example of a CPA is shown in Fig. \ref{fig2}d.

\begin{figure}[h]
\includegraphics[width=9cm,height=6cm,clip]{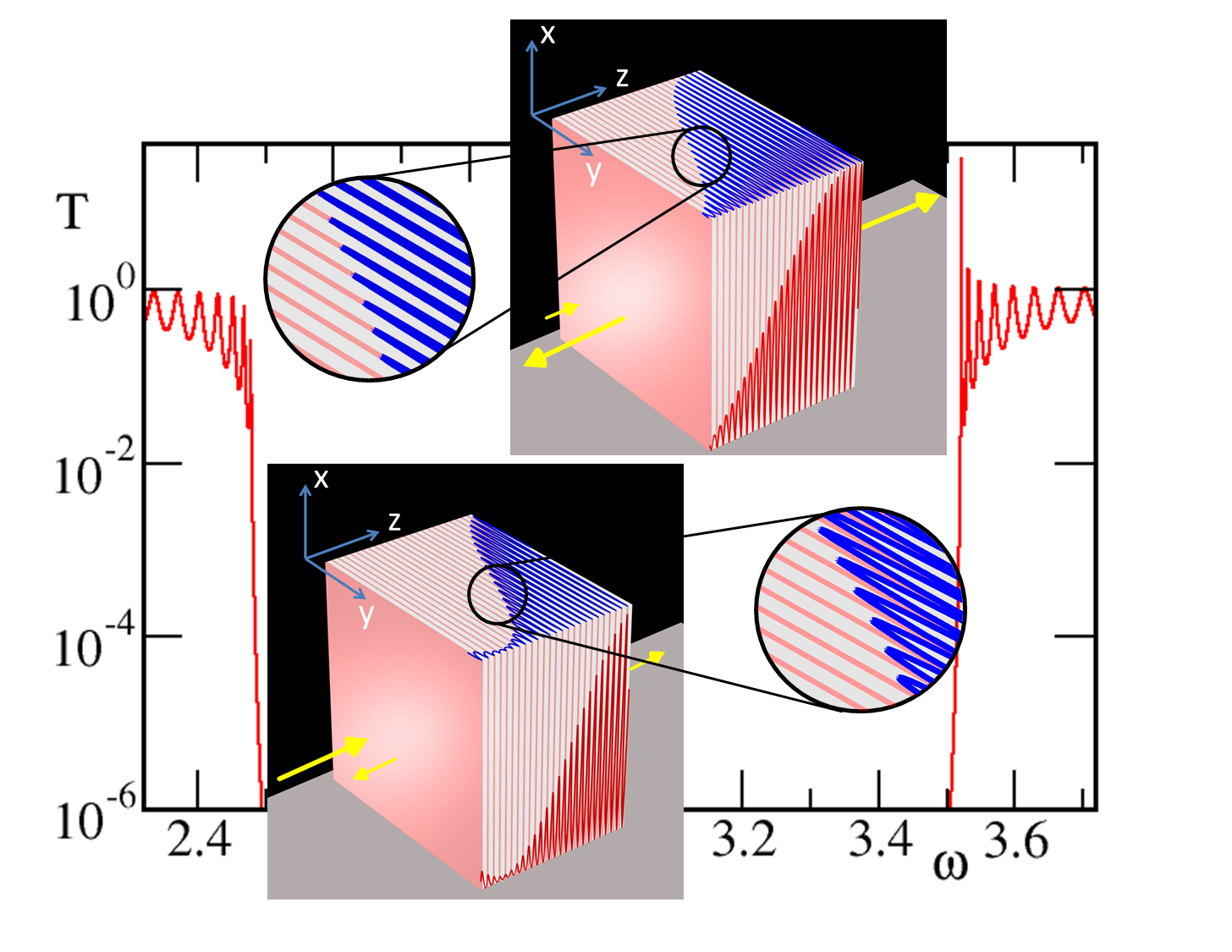}
 \caption{
   \label{fig3}
(color online) The electric $E(z)$ (blue line at y-z plane) and the magnetic $H(z)$ (red line at the x-z plane) fields of two representative 
FP modes which reside at the upper (left inset) 
and lower (right inset) band edges. The grating consists of $50$ bilayers (orange indicates a PP-layer and white a lossless passive layer). 
In the insets we show the field profiles for the first half of the structure, i.e. $25$ layers. Left (right) inset correspond to the FP resonance 
at the upper (lower) edge of the lower (upper) band. The constituents of the PP-layers are $\epsilon_p=4-0.001i, \mu_p=\epsilon_p^*$ 
with a width layer $d_p=0.1309 \mu m$. The lossless passive layers have constituents $\epsilon=3$ and $\mu=1$ and width $d=
0.3023 \mu m$. The transmittance ${\cal T}(\omega)$, of the grating is shown in the main panel.
}
\end{figure}

The above super/sub-unitary scattering features become more pronounced when one of the composite layers of the Bragg grating 
is substituted by a pseudo-passive medium. Specifically, the Fabry-Perrot (FP) resonances which are closer to the band edges will
correspond to photons with small group velocities. Therefore each individual photon at this frequency resides in the PP-layers for 
a long time. This long time interaction with the PP-medium leads to strong amplification/attenuation features. The magnitude of 
the enhancement/suppression is a growing function of the total thickness of the stack, as opposed to the case of a uniform pseudo
-passive slab. In Fig. \ref{fig3} we show a typical transmission spectrum of such composite structure.

In the particular case of a FP resonance close to a photonic band edge of a BG, consisting of two layers A and B, the field inside the 
periodic structure is a superposition of the forward and backward Bloch waves with Bloch wave number close to either the middle of 
the Brillouin zone, or to its boundary. In either case, the middle of each individual layer coincides with the node or the antinode of the 
respective standing wave inside this particular uniform layer. The nodes (antinodes) of oscillating electric field coincide with the antinodes 
(nodes) of oscillating magnetic field and vice versa. 

At the same time, if the lower edge of a photonic band gap has electric field node (magnetic field antinode) in the middle of each A-layer, 
then the upper edge of the same photonic band gap has electric field antinode (magnetic field node) in the middle of each A-layer \cite{Jbook}. 
As a consequence, Fabry-Perrot resonances close to neighboring photonic band edges will have different dominating components - electric 
or magnetic - of the oscillating field inside a particular layer.

In our example in Fig. \ref{fig3} the A-layers consists of a PP-medium with negative $\epsilon''$ and positive $\mu''$, which means that 
the Fabry-Perrot resonances with the dominant electric field component in the A-layers will be enhanced, while the resonances with 
dominant magnetic field component in the A-layers will be suppressed. This is exactly what we see in Fig. \ref{fig3}, where the resonant 
transmission at every other photonic band edge is enhanced (suppressed). 

In conclusion, we have introduced a class of pseudo-passive metamaterials for which the losses are balanced by gain in a way that 
the index of refraction is real and uniform throughout the medium. Although the light propagation in an unbounded pseudo-passive 
medium has the same characteristics as in a passive lossless medium with the same index of refraction, their scattering properties 
differ dramatically. When such media are incorporated in a photonic structure, it can lead to super/sub-unity transmittance/reflectance 
indicating strong absorption and amplification mechanisms for the total structure. In many occasions these effects can turn to a 
coherent perfect absorption of incident waves or to lasing instabilities. It will be interesting to investigate the realization of these  
phenomena under vectorial conditions and in higher dimensions.

\textit{Acknowledgment - } This work was partially supported by an AFOSR MURI grant FA9550-14-1-0037, by a LRIR-09RY04COR, 
and by a NSF ECCS-1128571 grant.



\begin{thebibliography}{99}
\bibitem{note1} In this case the super/sub unitary values of transmittance and reflectance can be achieved 
for even smaller values of $\epsilon^{\prime \prime}\leq 10^{-3}$ since now the photons can interact for larger times with the PP-
media which are distributed all over the grating.
\end{thebibliography}
\end{document}